# Resonant inelastic soft-x-ray scattering spectra at the $N_{1s}$ and $C_{1s}$ edges of poly(pyridine-2,5-diyl)

M. Magnuson[1], L. Yang[2], J.- H. Guo[1], C. Såthe[1], A. Agui[1], J. Nordgren[1], Y. Luo[3],
H. Ågren[2], N. Johansson[2], W. R. Salaneck[2], L. E. Horsburgh[4] and A. P. Monkman[4]

[1]*Department of Physics, Uppsala University, Box 530, S-75121 Uppsala, Sweden*
[2]*Institute of Physics and Measurement Technology, Linköping University, Box 530, S-58183 Linköping, Sweden*
[3]*FYSIKUM, University of Stockholm, Box 6730, S-113 85 Stockholm, Sweden*
[4]*Department of Physics, University of Durham, South Road, Durham DH1 2UE, England*

**Abstract**
Resonant inelastic scattering measurements of poly(pyridine-2,5-diyl) have been performed at the $N_{1s}$ and $C_{1s}$ edges using synchrotron radiation. For comparison, molecular orbital calculations of the spectra have been carried out with the repeat unit as a model molecule of the polymer chain. The resonant emission spectra show depletion of the p electron bands which is consistent with symmetry selection and momentum conservation rules. The depletion is most obvious in the resonant inelastic scattering spectra of carbon while the nitrogen spectra are dominated by lone pair *n* orbital emission of s symmetry and are less excitation energy dependent. By comparing the measurements to calculations an isomeric dependence of the resonant spectra is found giving preference to two of the four possible isomers in the polymer.

## Introduction

Conjugated polymers have been the subject of much interest owing to their unique electronic properties which can be technically exploited e.g., as doping induced electrical conductors and light emitting diodes [1]. Detailed experimental studies of the uppermost p-orbital levels at the valence band edges of these polymers are important to gain an understanding of their properties. Such studies have been carried out by many techniques including photoelectron spectroscopy using photon excitation in both the x-ray and ultraviolet wavelength regimes.

X-ray emission (XE) spectroscopy provides a useful technique for studying conjugated polymers but has yet not been exploited much. XE provides electronic structure information in terms of local contributions to the Bloch or molecular orbitals (MO's), since the x-ray processes can be described by local dipole selection rules. The method is atomic element specific and also angular momentum and symmetry selective at high resolution. However, the relatively low fluorescence yield and instrument efficiencies associated with XE in the sub keV energy region lead to considerable demands. An intense synchrotron radiation (SR) excitation source is therefore required which has earlier limited the experimental activity of studying the behavior for radiative emission spectroscopy of oligomers and polymers.

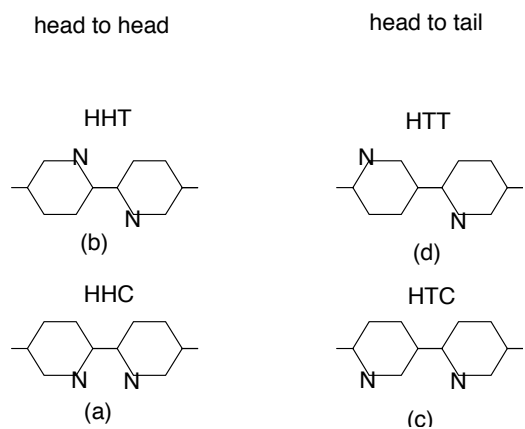

**Figure 1:** (panels a, b, c and d) The isomeric head-to-tail (H-T) and head-to-head (H-H) geometries of PPy. The extra letter, T and C (*trans* and *cis*) denotes the relative positions between adjacent nitrogen atoms

Non-resonant XE spectra are obtained when the energy of the incident photons exceed far above the core ionization threshold. In this case the XE spectral profile is practically independent (besides x-ray satellites) of the excitation energy and has been often described using a two-step model with the emission step decoupled from the excitation step. On the other hand, when the excitation energy is tuned at resonances below or close to a core ionization threshold, the spectral distribution is strongly dependent on the excitation energy. The description in the resonant case must therefore switch from a two-step to a one-step model with the excitation and emission transitions treated as a single scattering event in resonant inelastic x-ray scattering (RIXS).

In a recent work we used a set of poly(*p*-phenylenevinylene)s; PPV, PMPV and PDPV to demonstrate the feasability of studying the electronic structure of conjugated polymers by means of resonant and non-resonant XE measurements with monochromatic SR excitation [2]. It is of great interest to find out how the resonant and non-resonant spectra show up in more complicated hetero-compounds, and separately measure the energy bands from the different atomic elements involved. In the present work we analyze for this purpose, resonant XE spectra of poly(pyridine-2,5-diyl) (PPy) which is an aza-substituted poly(*p*-phenylene). The analysis based on *ab initio* canonical Hartree-Fock theory indicate isomeric dependence of the RIXS spectra [3]. The resonant XE spectra show that the p electron bands dissappear in the spectra due to symmetry selection and momentum conservation rules.

**Experiment**
The experiments were carried out at beamline 7.0 at the *Advance Light Source* (ALS) at the Lawrence Berkeley National Laboratory. This undulator beamline includes a spherical-grating monochromator and provides linearly polarized SR of high resolution and high brightness.
The XE spectra were recorded using a high-resolution grazing-incidence x-ray fluorescence spectrometer [4]. During the XE measurements, the resolution of the beamline monochromator was 0.25 eV and 0.40 eV for the carbon and nitrogen edges, respectively. The x-ray fluorescence spectrometer had a resolution of 0.30 eV and 0.65 eV, for the carbon and nitrogen measurements. The sample was oriented so that the incidence angle of the photons was 20 degrees with respect to the surface plane. During the data collection, the samples were scanned (moved every 30 seconds) in the photon beam to avoid photon-induced decomposition of the polymers. The base pressure in the experimental chamber was 4 x 10$^{-9}$ Torr during the measurements.

## Calculations

Figure 1 illustrates head-to-head (HH) and head-to-tail (HT) repeat units of the four different isomers possible in the PPy conjugated polymer. An extra letter, T or C (*trans* or *cis*), denotes the relative positions between adjacent nitrogen atoms. The calculations were carried out by taking a repeat unit (dimer) as a model molecule of the PPy polymer which includes two polymer rings. The restriction to one single (dimer) unit is motivated by the fact that PPy has relatively flat bands. A few test calculations on multiple-unit spectra did not give significant differences. The geometries of the model molecules were obtained by using the AM1 Hamiltonian [5] in the MOPAC program for a model molecule with four pyridine rings. Calculations of the pyridine molecule were also carried out for comparison. The direct self-consistent field (SCF) program DISCO [6] has been employed for calculations of the orbital energies and dipole transition moments.

## Results

Figure 2 shows a resonant XE spectrum of PPy excited at 398.8 eV photon energy at the $N_{1s}$ threshold. The corresponding calculated spectra of the four different isomers and the pyridine molecule are presented below. The spectrum basically consists of two main parts. A strong elastic (recombination) peak at 398.8 eV has an energy position equivalent to the excitation energy. The other part constitutes the inelastic scattering. In PPy, five features (labeled A-E) can be identified in the inelastic scattering part of the spectrum. Peak A corresponds to p-electron and n electron states (1a$_2$ and 11a$_1$(n) in pyridine), peak B is due to both p and s, with s dominating, while features C, D, and E, are all related to s electronic states. The five bands clearly show up in the calculation of the pyridine molecule as well as for the four different isomers of PPy which were obtained by taking the repeat-unit as a model molecule. In the polymer case, the bands are broader due to a large number of transitions involved in each band. It is also interesting to note that the bands A-E observed in these RIXS spectra correspond well in energy with the five bands observed in ultraviolet photoelectron spectra (UPS) [7]. However, the intensities are different owing to the different nature of the transition moments between the spectroscopies.

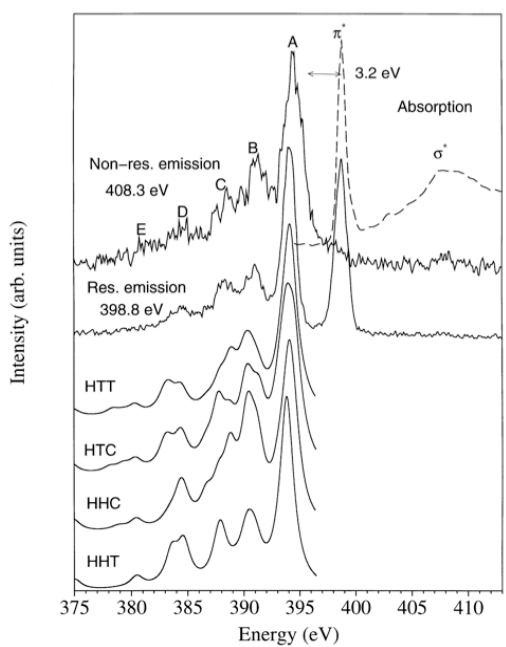

**Figure 2:** Experimental and calculated resonant $N_{1s}$ XE spectra of PPy.

In Fig. 2 the relative intensity of peak B is about 73% as high as that of peak A, for the HHC isomeric form, whereas for the other three isomers the ratio is about 45%. Some differences in the intensity distribution of the C-bands are also predicted by the calculations. The comparison with the experimental spectrum appear to favour the HTT and the HHT isomeric forms and disfavour the HHC form.

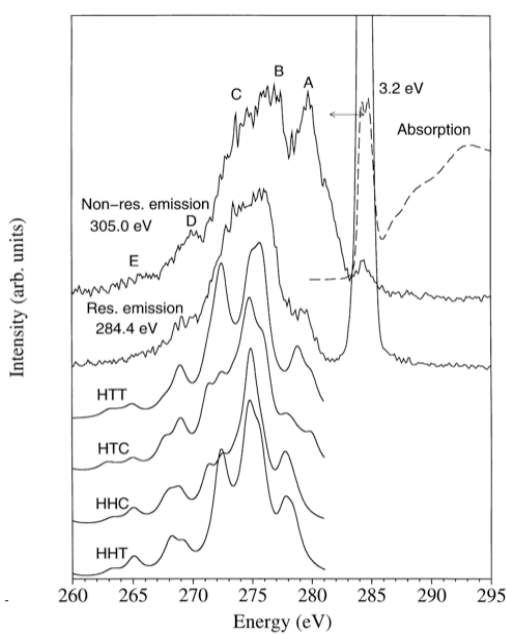

**Figure 3:** Experimental and calculated resonant $C_{1s}$ XE spectra of PPy.

Figure 3 shows a resonant XE spectrum of PPy excited at 284.4 eV photon energy at the $C_{1s}$ threshold. In this case the elastic (recombination) peak is stronger due to a different localization character of the emitting levels. The carbon spectrum obviously map the same final states of the inelastic scattering part as the nitrogen spectrum but with different transition moments owing to the different intermediate states. Thus, the carbon spectrum show a similar relative peak position of the bands while the intensity distribution is different. In both cases the spectra show strong elastic (recombination) peaks. The calculated $C_{1s}$ spectrum of the pyridine molecule shows a similar band structure as the $N_{1s}$ spectrum, but with a different intensity distribution among the bands, reflecting the localization of the core-hole orbitals. The most conspicuous difference between the nitrogen and carbon spectra is the intensity of peak A. It becomes less intense in the resonant $C_{1s}$ spectrum, since the *n*-lone pair MO's have only little overlap with the localized $C_{1s}$ core orbital. On the other hand, the MO's of band B have larger 2p contributions from the carbon atoms and so a larger dipole overlap with the $C_{1s}$ core orbitals than with the $N_{1s}$ core orbitals is expected. For the inner MO's, such as band E, the intensities are weak in both the $C_{1s}$ and $N_{1s}$ spectra due to the larger 2s character of the MO's and also due to a breakdown of the MO approximation accompanying correlation state splittings [8]. The calculated resonant spectra for the different isomers show an isomeric dependence. As in the nitrogen case the best agreement between experiment and theory is obtained for the HTT and the HHT isomeric forms. It has also been found that the calculated density-of-states distribution in comparison to the measured UPS spectrum of PPy by Miyamae *et al.* [7], gives some preference to the head-to-tail isomeric form.

## Discussion

For p-electron systems, the momentum conservation leads to depletion of emission from the p-levels which has previously been observed in RIXS spectra of Benzene [9] and aniline [10]. A similar effect has also been found in conjugated polymer compounds such as PPV, PMPV and PDPV [2]. In both cases a similar depletion of the p-levels could be obtained theoretically by a full symmetry and interference analysis of the resonant process using the repeat unit as a model. The depletion has about the same strength in the PPy polymer as for the PPV compounds although a smaller effect would be expected for PPy due to the stronger chemical shifts of the core-excited states. However, a corresponding depletion is not observed in the resonant nitrogen spectrum. This can be explained by the fact that the strong high-energy band in the nitrogen spectrum is dominated by the lone-pair $n$ orbitals localized on the nitrogen sites, which have s symmetry. It is also relevant to consider the different localization characters of the emitting levels in the interpretation of the resonant spectra.

## Summary

Resonant inelastic soft-x-ray scattering spectra of poly(pyridine-2,5-diyl) have been measured with monochromatic synchrotron radiation. The spectra exhibit peak structures far into the valence energy region. The relative energy positions of the emission bands are similar in the carbon and nitrogen spectra i.e., the same types of final states are involved but different relative intensities are observed since the localization properties of the energy bands are different. In particular, the lone-pair $n$ levels emphasize the high energy part of the nitrogen spectrum compared to the carbon spectrum. The most conspicuous difference is the partial depletion of the p band in the carbon case which is also expected from the theory of resonant inelastic x-ray scattering spectra of p-electron systems. The comparison to canonical Hartree-Fock calculations indicate isomeric dependence of the resonant x-ray emission spectra, and gives preference for two of the four isomers contained in the conjugated polymer.

## Acknowledgements

This work was supported by the Swedish Natural Science Research Council (NFR), the Swedish Research Council for Engineering Sciences (TFR), the Göran Gustavsson Foundation for Research in Natural Sciences and Medicine and the Swedish Institute (SI). The experimental work at the ALS, Lawrence Berkeley National Laboratory was supported by the Office of Energy Research, Office of Basic Energy Science, Material Science Division of the U. S. Department of Energy, under contract No. DE-AC03-76SF00098.